\documentclass[12pt]{article}

\usepackage{amssymb}
\usepackage{amsmath}

\newcommand{\eps}{\varepsilon}
\newcommand{\la}{\lambda}
\newcommand{\prt}{\partial}
\newcommand{\K}{{\mathrm{K}} }
\newcommand{\E}{{\mathrm{E}} }

\newcommand{\sn}{{\mathrm{sn}} }

\begin{document}


\title{On Whitham theory for perturbed integrable equations}

\author{A.M. Kamchatnov\thanks{E-mail address: kamch@isan.troitsk.ru}\\
Institute of Spectroscopy, Russian Academy of Sciences,\\
Troitsk, Moscow Region, 142190, Russia\\
and \\
 Departamento de F\'{\i}sica e Centro de F\'{\i}sica de
Mat\'eria Condensada,\\ Universidade de Lisboa,  Av.~Prof.~Gama Pinto 2,\\
Lisbon 1649-003, Portugal }

\maketitle

\begin{abstract}
Whitham theory of modulations is developed for periodic waves
described by nonlinear wave equations integrable by the inverse
scattering transform method associated with $2\times2$ matrix or
second order scalar spectral problems. The theory is illustrated
by derivation of the Whitham equations for perturbed Korteweg-de
Vries equation and nonlinear Schr\"odinger equation with linear
damping.
\end{abstract}




\section{Introduction}

The theory of perturbations of soliton solutions of integrable
equations is well developed and has found many applications to
various physical problems (see, e.g. \cite{KM89}). On the
contrary, perturbations of wave trains did not attract much
attention, although there are many problems which should be
treated in framework of such perturbational approach. As the most
evident applications we can mention the theory of dissipationless
shock waves in nonlinear media described by the Korteweg-de Vries
(KdV) \cite{GP73}, nonlinear Schr\"odinger (NLS) \cite{KKU02},
derivative NLS \cite{GKE92}, Kaup-Boussinesq (KB) \cite{EGP01},
and many other equations. As other examples one can also mention
fluxon trains in long Josephson junctions \cite{ustinov98}
described by the sine-Gordon equation and soliton trains in
Bose-Einstein condensate confined by magnetic traps
\cite{nature02} and described by the NLS equation with external
potential. In all these cases unperturbed nonlinear equations have
periodic solutions and typical problems can be often reduced to
finding slow evolution of parameters of the nonlinear wave due to
its initial modulation and small perturbation terms in the
governing wave equations.

The general approach to treatment of such problems was suggested long ago
by Whitham \cite{whitham65,whitham74} and it is based on two scales of
time---fast oscillations of the wave and slow evolution of its parameters.
Then equations for slow evolution of the parameters can be averaged over
fast oscillations what leads to the system of the first order partial
differential equations called Whitham equations. Whitham discovered that
in the case of (unperturbed) KdV equation the arising in this way system
can be drastically simplified by introduction of proper parameterization
of the periodic solution when Whitham equations take the so-called
diagonal form for slowly varying parameters called Riemann invariants.
Later it was realized \cite{FFML80} that this property of the KdV
equation is connected with its complete integrability by the inverse
scattering transform (IST) method and just this method yields in a
natural way the periodic solutions of the KdV equation parameterized
by the Riemann invariants. In the fundamental paper \cite{FFML80} Whitham
equations were derived for the general case of quasi-periodic solutions
of the KdV equation. After that this method was applied to other
integrable equations \cite{FML83,FL86,pavlov87} and used for description
of dissipationless shock waves in modulationally stable systems
\cite{GKE92a,tian93,KKU02} and formation of soliton trains at sharp
fronts of pulses in modulationally unstable systems (see \cite{kamch97}
and references therein).

The Whitham theory approach implies that the process under
consideration is considered at asymptotically large periods of
time much greater than the period of wave oscillations. However,
it is clear that small perturbations neglected in completely
integrable approximations can make considerable contribution into
long-time evolution. Therefore it is necessary to develop
perturbation theory in the Whitham approach and the first step
should be development of the method of derivation of perturbed
Whitham equations. The method suggested in \cite{FML84} for
perturbed KdV equation was quite complicated and has not
found many applications. Therefore the authors of
refs.~\cite{GP87,AN87,MG95} used the classical
Whitham approach for consideration of influence of weak
dissipation on long-time evolution of the KdV wave trains. In a
similar way, the Whitham equations for two-phase quasi-periodic
solutions of perturbed sine-Gordon equation obtained in
\cite{BEK98} did not have effective enough form and were
considered in a very special case only.

The aim of this paper is to develop simple and general enough
method of obtaining Whitham equations for modulated wave train
solutions of perturbed nonlinear equations. The method can be
applied to integrable nonlinear wave equations associated, in
framework of the IST method, with second order scalar spectral
problem or equivalent to it $2\times 2$ matrix problem
(Ablowitz-Kaup-Newell-Segur scheme \cite{AKNS} and its generalizations).
The general formulae obtained in the next section are then
illustrated by applications to the perturbed KdV equation (where known
results are reproduced)
and to the NLS equation with weak linear damping. These examples
show that our approach is quite effective and applicable to many
nonlinear wave equations of physical importance with various
types of perturbations.

\section{Perturbed Whitham equations}

Many physically important equations can be solved, in principle,
analytically by the IST method in which they are represented as
compatibility conditions of two linear systems with spectral
parameter (see, e.g. \cite{kamch2000}). We shall confine ourselves
to the cases when these two systems can be represented in one of
two equivalent forms. The $2\times 2$ matrix form is given by
\begin{equation}\label{eq1}
\begin{array}{c}
\Psi_x =\mathbb{U}\Psi,\qquad \Psi_t=\mathbb{V}\Psi,\\
\Psi= \left( \begin{array}{c} \psi_1 \\ \psi_2
\end{array}\right),\quad
\mathbb{U}= \left( \begin{array}{cc}
F & G\\
H & -F
\end{array}\right),\quad
\mathbb{V}= \left(\begin{array}{cc}
A & B\\
C & -A
\end{array}\right),
\end{array}
\end{equation}
where the matrix elements depend on as field variables
$u_m(x,t)$ of the equations under consideration, so the spectral
parameter $\la$. For example, in the case of the NLS equation,
\begin{equation}\label{1-1}
  \mathrm{i}u_t+u_{xx}\pm 2|u|^2u=0,
\end{equation}
these matrix elements correspond to the Zakharov-Shabat
spectral problem \cite{ZS73} and are equal to
\begin{equation}\label{1-2}
\begin{array}{l}
  F=-\mathrm{i}\la,\quad G=\mathrm{i}u,\quad H=\pm \mathrm{i}|u|^2; \\
  A=-2\mathrm{i}\la^2\pm \mathrm{i}|u|^2,\quad B=2\mathrm{i}u\la-u_x,\quad
  C=\pm 2\mathrm{i}u^*\la\pm u_x^*,
  \end{array}
\end{equation}
and there are two field variables, $u(x,t)$ and its complex conjugate
$u^*(x,t)$.
The compatibility condition $\Psi_{xt}=\Psi_{tx}$ of the systems
(\ref{eq1}) yields at once the evolution equations in general form
\begin{equation}\label{eq2}
\begin{array}{l}
  F_t-A_x+CG-BH=0,\\
  G_t-B_x+2(BF-AG)=0,\\
  H_t-C_x+2(AH-CF)=0.
  \end{array}
\end{equation}
Taking as an example again the NLS equation, it is easy to see that
if we substitute (\ref{1-2}) into (\ref{eq2}) and demand that these
equations must fulfil for any value of the spectral parameter $\la$,
then we find the NLS equation and its complex conjugate,
\begin{equation}\label{2-1}
  u_t=\mathrm{i}u_{xx}\mp 2\mathrm{i}|u|^2u, \quad
  u_t^*=-\mathrm{i}u_{xx}^*\pm 2\mathrm{i}|u|^2u^*.
\end{equation}
In general case (\ref{eq1}) we shall write symbolically the
equations in terms of the field variables $u_m(x,t)$ as
\begin{equation}\label{eq3}
  u_{m,t}=K_m(u_n, u_{n,x},\ldots ),\qquad m,n =1,\ldots, N,
\end{equation}
Many physically important
nonlinear equations are associated with linear systems (\ref{eq1}).

Another common form of linear equations associated with nonlinear
integrable equations can be written as
\begin{equation}\label{eq4}
  \psi_{xx}=\mathcal{A}\psi,
\end{equation}
\begin{equation}\label{eq5}
   \psi_t=-\frac12 \mathcal{B}_x\psi+ \mathcal{B}\psi_x,
\end{equation}
where again the coefficients $\mathcal{A}$ and $\mathcal{B}$ depend on the
field variables and the spectral parameter $\la$.
For well-known example of the KdV equation
\begin{equation}\label{3-1}
  u_t+6uu_x+u_{xxx}=0
\end{equation}
there is only one field variable $u(x,t)$ and
\begin{equation}\label{3-2}
  \mathcal{A}=-(u+\la),\qquad \mathcal{B}=4\la-2u.
\end{equation}
The compatibility condition $(\psi_{xx})_t=(\psi_t)_{xx}$
of the equations (\ref{eq4}),(\ref{eq5}) yields
\begin{equation}\label{eq6}
\mathcal{A}_t-2\mathcal{B}_x\mathcal{A}-\mathcal{B}\mathcal{A}_x+
\frac12\mathcal{B}_{xxx}=0.
\end{equation}
After substitution of (\ref{3-2}) this equation yields the KdV equation
(\ref{3-1}), and in the general case (\ref{eq4}),(\ref{eq5})
the resulting evolution equations can be represented again in the form
(\ref{eq3}). Besides the KdV equation,
the representation (\ref{eq4}),(\ref{eq5}) is convenient for treatment of the
KB system \cite{Kaup} and some other equations. As was
shown in \cite{KK02}, the matrix system (\ref{eq1}) can be
transformed to scalar form (\ref{eq4}),(\ref{eq5}) by the formulae
\begin{equation}\label{eq7}
\mathcal{A}=(F-G_x/2G)^2+GH+(F-G_x/2G)_x,
\end{equation}
\begin{equation}\label{eq8}
\mathcal{B}=B/G,
\end{equation}
so that equations (\ref{eq2}) reduce to (\ref{eq6}). Thus, these two
forms of linear spectral problems are equivalent to each other and
both can be used for treatment of the respective nonlinear equations.
The scalar form (\ref{eq4}),(\ref{eq5}) is more convenient for
derivation of perturbed Whitham equations and therefore it will be
used in what follows.

We assume that unperturbed wave equations (\ref{eq3}) have periodic
solutions which can be obtained in framework of the IST method in
the following way (see, e.g. \cite{kamch2000}). We take two basis
solutions $\psi^+$ and $\psi^-$ of the second order differential
equation (\ref{eq4}) and construct from them the so-called `squared
basis function'
\begin{equation}\label{eq9}
  g=\psi^+\psi^-.
\end{equation}
It is easy to show that it satisfies the equation
\begin{equation}\label{eq10}
 g_{xxx}-2\mathcal{A}_x{g}-4\mathcal{A}{g}_x=0,
\end{equation}
which upon multiplication by $g/2$ can be integrated once to give
\begin{equation}\label{eq11}
\frac12{g}{g}_{xx}-\frac14{g}_x^2-\mathcal{A}{g}^2
={P}(\la),
\end{equation}
where the integration constant denoted by $P(\la)$ can depend on the
spectral parameter $\la$. In the finite-gap integration method (see, e.g.
\cite{ZMNP}) it is shown that periodic solutions are distinguished by
the condition that $P(\la)$ be a polynomial in $\la$ . Then $g$ as a
function of $\la$ should also be a polynomial in $\la$ with coefficients
expressed in terms of $u_m(x,t)$, and, hence, $u_m(x,t)$ can be found
from equations obtained by equating coefficients of $\la^s$ in
Eq.~(\ref{eq11}) (we suppose here that $\mathcal{A}$ is a rational
function of $\la$ in accordance with majority of applications). The
dependence of $g$, and hence of its coefficients, on time $t$ is
obtained from the equation
\begin{equation}\label{eq12}
  {g}_t=\mathcal{B}{g}_x-\mathcal{B}_x{g}
\end{equation}
which follows from Eqs.~(\ref{eq5}) and (\ref{eq9}). As a result,
$u_m(x,t)$ are expressed as (quasi)periodic functions of $x$ and
$t$ parameterized by the zeroes $\la_i$ of the polynomial
$P(\la)=\prod_{i=1}^M(\la-\la_i)$,
\begin{equation}\label{eq13}
  u_m(x,t)=u_m(x,t;\la_1,\ldots,\la_M), \quad m=1,\ldots, N.
\end{equation}

In strictly periodic solutions the parameters $\la_i$ are constant
by definition. But we can slightly modulate this periodic wave by
making $\la_i$ slow functions of $x$ and $t$ which change little
in one wavelength and one period. Whitham showed
\cite{whitham65,whitham74} (for particular case of one-phase
periodic solution of the KdV equation when this solution is
parameterized by three zeroes $\la_1,\,\la_2,\,\la_3$) that substitution of
\begin{equation}\label{eq14}
  u_m(x,t)=u_m(x,t;\la_1(x,t),\ldots,\la_M(x,t)), \quad m=1,\ldots, N,
\end{equation}
into conservation laws of Eqs.~(\ref{eq3}) followed by averaging of
these conservation laws over fast oscillations of the wave
yields for slow variables $\la_i(x,t)$
the so-called Whitham modulation equations
\begin{equation}\label{eq15}
  \frac{\prt\la_i}{\prt t}+v_i(\la_j)\frac{\prt\la_i}{\prt x}=0,
  \quad i,j=1,\ldots,M,
\end{equation}
which govern slow evolution of modulated periodic solutions of
unperturbed equations (\ref{eq3}). This approach is well developed
and finds a number of applications (see, e.g. \cite{kamch2000}).

Now, we want to take into account perturbations in the wave equations
(\ref{eq3}),
\begin{equation}\label{eq16}
  u_{m,t}=K_m(u_n, u_{n,x},\ldots )
  +R_m(x,t,u_n, u_{n,x},\ldots ),\qquad m,n =1,\ldots, N,
\end{equation}
where the perturbation terms $R_m$ can be slow functions of $x$ and
$t$ and can also depend on the field variables $u_n$ and their space
derivatives. The perturbation terms can have any nature, e.g., they can
arise due to external potentials or small physical effects neglected
in derivation of unperturbed equations (\ref{eq3}). Obviously, the
functions (\ref{eq13}) are not the solutions of the perturbed
equations (\ref{eq16}) anymore, but again, as in the case of slow
modulation, we can describe effects of perturbations as slow change
of the parameters $\la_i,$ $i=1,\ldots,M$. Then after averaging over
fast oscillations of the wave we have to obtain additional terms in
the Whitham equations (\ref{eq15}) which represent effects of
perturbations. Our aim is to derive such generalization of the
Whitham equations.

As was indicated by Whitham \cite{whitham65,whitham74}, the
most convenient way of obtaining the modulation equations is to
average the conservation laws. Completely integrable equations
have an infinite sequence of conservation laws and averaging of
their generating function yields for periodic solutions of equations
(\ref{eq3}) associated with the spectral problems (\ref{eq1}) or
(\ref{eq4}),(\ref{eq5}) the modulation Whitham equations
directly in diagonal Riemann form \cite{FFML80}. Therefore the
starting point of our analysis should be the formula for
generating function of conservation laws for equations (\ref{eq6})
described by the IST scheme (\ref{eq4}),(\ref{eq5}). As was
mentioned above, we suppose that $ \mathcal{A}$ is a rational function
of $\la$. Then it can be chosen so that
\begin{equation}\label{eq17}
  \left. \mathcal{A}\right|_{\la\to\infty}\longrightarrow \sigma\la^r,
\end{equation}
where $\sigma=\pm 1$ and $r$ is an integer number. With such a
normalization of $ \mathcal{A},$ we can rewrite Eq.~(\ref{eq11})
in the form
\begin{equation}\label{eq18}
\frac12{g}{g}_{xx}-\frac14{g}_x^2-\mathcal{A}{g}^2
=-\sigma {P}(\la),
\end{equation}
where $P$ and $g$ are polynomials in $\la$,
\begin{equation}\label{eq19}
  P=\prod_{i=1}^M(\la-\la_i),\quad
  g=\prod_{i=1}^{{\rm deg}g}(\la-\mu_i(x,t)),
\end{equation}
such that their degrees are connected by the relation
$M=r+2\,{\rm deg}\,g$. If we introduce new function
\begin{equation}\label{eq20}
  \tilde{g}=\frac{\la^{r/2}}{\sqrt{P(\la)}}g,
\end{equation}
then the identity (\ref{eq18}) takes the form
\begin{equation}\label{eq21}
\frac1{2\la^r}\tilde{g}\tilde{g}_{xx}-\frac1{4\la^r}\tilde{g}_x^2
-\tilde{\mathcal{A}}\tilde{g}^2=-\sigma,
\end{equation}
where
\begin{equation}\label{eq22}
  \tilde{\mathcal{A}}= \mathcal{A}/\la^r,
\end{equation}
and
\begin{equation}\label{eq23}
  \left.\tilde{g}\right|_{\la\to\infty}=1.
\end{equation}
Thus, we can expand $\tilde{g}$ in inverse powers of $\la$,
\begin{equation}\label{eq24}
  \tilde{g}=\sum_{n=0}^{\infty}\frac{g_n}{\la^n}, \quad g_0=1,
\end{equation}
and the coefficients $g_n$ can be calculated in recurrent way by
substitution of (\ref{eq24}) into (\ref{eq21}). The examples of this
procedure will be given below.

Now, from (\ref{eq12}) and (\ref{eq20}) we find that $\tilde{g}$
satisfies the relation
\begin{equation}\label{eq25}
  \left(\frac1{\tilde{g}}\right)_t=
  \left(\frac{\mathcal{B}}{\tilde{g}}\right)_x
\end{equation}
which can be considered as a generating function of conservation laws.
Indeed, the coefficients of the expansion
\begin{equation}\label{eq26}
  \frac1{\tilde{g}}=\sum_{n=0}^{\infty}\frac{g_{-n}}{\la^n}, \quad g_0=1,
\end{equation}
can be expressed in terms of the already known coefficients $g_n$ with the use of
trivial identity $\tilde{g}\cdot(1/\tilde{g})=1$, and then these
coefficients $g_{-n}$ serve as the densities of the conservation laws and
the coefficients of similar expansion of $ \mathcal{B}/\tilde{g}$ in
inverse powers of $\la$ serve as the corresponding fluxes,
\begin{equation}\label{9-1}
  \left(g_{-n}\right)_t=\left[(\mathcal{B}/\tilde{g})_{-n}\right]_x,
\end{equation}
where $(\mathcal{B}/\tilde{g})_{-n}$ denotes the coefficients in the
expansion
\begin{equation}\label{9-2}
  \frac{\mathcal{B}}{\tilde{g}}=\sum_{n=0}^{\infty}
  \frac{(\mathcal{B}/\tilde{g})_{-n}}{\la^n}.
\end{equation}

The functions $\tilde{g}$ and $1/\tilde{g}$ satisfy important
identities
\begin{equation}\label{eq38}
  \frac{ \widehat{\delta} }{ \delta u_m} \left(\frac1{\tilde{g}}\right)=
  \frac{\sigma}{2\la^{r}}
  \sum_{l=0}^{A_m}(-1)^l\frac{\prt^l}{\prt x^l}\left(\frac{\prt\mathcal{A}}
  {\prt u_m^{(l)}}\tilde{g}\right), \quad m=1,\ldots, N,
\end{equation}
derived in the Appendix. Here ${ \widehat{\delta} }/{ \delta u_m}$
denotes the ``Euler derivative''
\begin{equation}\label{eq30}
\frac{ \widehat{\delta} }{ \delta u_m}=\frac{\prt}{\prt u_m}-
\frac{\prt}{\prt x}\frac{\prt}{\prt u_{m,x}}+
\frac{\prt^2}{\prt x^2}\frac{\prt}{\prt u_{m,xx}}-\ldots+
(-1)^{N_{mn}}\frac{\prt^{N_{mn}}}{\prt x^{N_{mn}}}\frac{\prt}{\prt u_{m}^{(N_{mn})}},
\end{equation}
when it is applied to $g_{-n}$, $N_{mn}$ is the order of the highest
derivative $u_m^{(N_{mn})}$ in $g_{-n}$, and $A_m$ is the order of the highest derivative
$u_m^{(A_m)}$ in $\mathcal{A}$.

Now let us turn to perturbed equations (\ref{eq16}) and calculate the
correction to the conservation law (\ref{9-1}) with the density
\begin{equation}\label{eq39}
  g_{-n}=g_{-n}(u_m,u_{m,x},\ldots, u_m^{(N_{mn})}).
\end{equation}
Evidently, we have
$$
\begin{array}{ll}
\left(g_{-n}\right)_t &= \sum_{m,k}g_{-n,mk}\prt_tu_m^{(k)}=
\sum_{m,k}g_{-n,mk}(\prt_x)^k\prt_tu_m\\
&=\sum_{m,k}g_{-n,mk}(\prt_x)^kK_m+\sum_{m,k}g_{-n,mk}(\prt_x)^kR_m,
\end{array}
$$
where $g_{-n,mk}$ denotes the derivative of $g_{-n}$ with respect to $u_m^{(k)}$.
The terms with $K_m$ must collect into divergence of the flux,
$\prt_x[(\mathcal{B}/\tilde{g})_{-n}]$, so that we obtain
\begin{equation}\label{eq40}
  \left(g_{-n}\right)_t=\left[(\mathcal{B}/\tilde{g})_{-n}\right]_x+
  \sum_{m,k}g_{-n,mk}(\prt_x)^kR_m.
\end{equation}
In what follows we shall average these conservation laws over some interval
$\Lambda$ of $x$ which is supposed to be much greater than the wavelength
$L$ and much less than the characteristic interval along which the
parameters of the wave change considerably. Therefore we can transform the
terms with $R_m$ by ``integration by parts'' and neglect the contribution
of boundaries because in the averaged form it is divided by $\Lambda$ and
goes to zero in asymptotic limit $\Lambda\gg L$ of very slow modulation.
As a result of this transformation we obtain the conservation law in the
following equivalent form:
$$
  \left(g_{-n}\right)_t=\left[\left(\frac{\mathcal{B}}{\tilde{g}}\right)_{-n}\right]_x+
  \sum_{m}\left(\frac{ \widehat{\delta} }{ \delta u_m}g_{-n}\right)R_m.
$$
After dividing this formula by $\la^n$ and summation over $n$ we arrive at
the generating function of perturbed conservation laws
\begin{equation}\label{eq42}
  \left(\frac1{\tilde{g}}\right)_t=
  \left(\frac{\mathcal{B}}{\tilde{g}}\right)_x+
  \sum_{m}\left(\frac{ \widehat{\delta} }{ \delta u_m}\frac1{\tilde{g}}\right)R_m.
\end{equation}
For further transformation we apply the identity (\ref{eq38}) and get
$$
    \left(\frac1{\tilde{g}}\right)_t-
  \left(\frac{\mathcal{B}}{\tilde{g}}\right)_x=
  \frac{\sigma}{2\la^{r}}
  \sum_{m,\,l}(-1)^l\frac{\prt^l}{\prt x^l}\left(\frac{\prt\mathcal{A}}
  {\prt u_m^{(l)}}\tilde{g}\right)R_m.
$$
Here we again can ``integrate by parts'' and neglect the boundary terms
to obtain
\begin{equation}\label{eq44}
    \left(\frac1{\tilde{g}}\right)_t-
  \left(\frac{\mathcal{B}}{\tilde{g}}\right)_x=
  \frac{\sigma}{2\la^{r}}
  \sum_{m,\,l}\frac{\prt\mathcal{A}}{\prt u_m^{(l)}}\frac{\prt^lR_m}
  {\prt x^l}\tilde{g}.
\end{equation}
To average this expression, it is convenient to return to the squared
basis function $g$ with the help of Eq.~(\ref{eq12}),
\begin{equation}\label{eq45}
    \left(\frac{\sqrt{P(\la)}}{{g}}\right)_t-
  \left(\sqrt{P(\la)}\,\frac{\mathcal{B}}{{g}}\right)_x=
  \frac{\sigma}{2}
  \sum_{m,\,l}\frac{\prt\mathcal{A}}{\prt u_m^{(l)}}\frac{\prt^lR_m}
  {\prt x^l}\frac{g}{\sqrt{P(\la)}}.
\end{equation}
Here the zeroes $\la_i$ of the polynomial $P(\la)$ are slow functions of
$x$ and $t$ which have to be differentiated with respect to $x$ and $t$
in the left hand side of Eq.~(\ref{eq45}). Then we obtain the terms with
$$
\frac1{\sqrt{\la-\la_i}}\frac{\prt\la_i}{\prt t}\quad {\rm and}\quad
\frac1{\sqrt{\la-\la_i}}\frac{\prt\la_i}{\prt x},
$$
which are singular at $\la\to\la_i$, and in the right hand side of
Eq.~(\ref{eq45}) we
already have terms with the same singular behaviour. Since the
perturbed conservation laws (\ref{eq45}) after averaging should be true at
any $\la$ including $\la_i$, we obtain
\begin{equation}\label{eq46}
  \frac{\prt\la_i}{\prt t}+v_i\frac{\prt\la_i}{\prt x}=
  \frac{\sigma}{2\langle1/g\rangle\prod_{j\neq i}(\la_i-\la_j)}
  \sum_{m=1}^N\sum_{l=0}^{A_m}\Big\langle\frac{\prt\mathcal{A}}
  {\prt u_m^{(l)}}\frac{\prt^lR_m}
  {\prt x^l}{g}\Big\rangle, \quad i=1,\ldots,M,
\end{equation}
where
\begin{equation}\label{eq47}
  v_i=-\frac{\langle \mathcal{B}/g\rangle}{\langle1/g\rangle},\quad
  i=1,\ldots,M,
\end{equation}
angle brackets denote averaging over proper interval of $x$, and everywhere
$\la$ is put equal to $\la_i$. Formula (\ref{eq47}) gives well-known
expressions for characteristic velocities in unperturbed Whitham equations.
The right hand side of (\ref{eq46}) describes the effects of perturbations.
Equations (\ref{eq46}) represent the result we wanted to obtain. The
averaged values of the expressions in angle brackets can be evaluated
with the use of formulae for periodic solutions of particular integrable
equations under consideration (see, e.g., \cite{FFML80,kamch94}).

Let us illustrate the obtained expressions for perturbed Whitham
equations with two examples.

\section{Examples}

We shall consider here simple examples of application of general formulae
(\ref{eq46}),(\ref{eq47}) to one-phase periodic solutions of the KdV and
NLS equations.

\subsection{Korteweg-de Vries equation}

In the case of the KdV equation
\begin{equation}\label{eq48}
  u_t+6uu_x+u_{xxx}=R
\end{equation}
we have expressions (\ref{3-2}) for $ \mathcal{A}$ and $\mathcal{B}$.
From (\ref{eq21}) we find the expansion
\begin{equation}\label{eq50}
  \tilde{g}=1-\frac{u/2}{\la}+\frac1{\la^2}\left(\frac38u^2+\frac18u_{xx}
  \right)-\frac1{\la^3}\left(\frac5{16}u^3+\frac5{32}u_x^2+
  \frac5{16}uu_{xx}+\frac1{32}u^{(IV)}\right)+\ldots,
\end{equation}
hence
\begin{equation}\label{eq51}
  \frac1{\tilde{g}}=1+\frac{u/2}{\la}-\frac1{\la^2}\left(\frac18u^2+\frac18u_{xx}
  \right)+\frac1{\la^3}\left(\frac1{16}u^3+\frac5{32}u_x^2+
  \frac3{16}uu_{xx}+\frac1{32}u^{(IV)}\right)+\ldots,
\end{equation}
and it is easy to check that these first terms of the expansions satisfy
the identity (\ref{eq38}), i.e.
\begin{equation}\label{eq52}
  \frac{ \widehat{\delta} }{ \delta u}\frac1{\tilde{g}}=
  \frac{\tilde{g}}{2\la}.
\end{equation}
The conservation laws following from Eq.~(\ref{9-1}) have well-known
form (see, e.g. \cite{kamch2000}).

The periodic solution (``cnoidal wave'') of unperturbed KdV equation has
the form (see, e.g. \cite{kamch2000})
\begin{equation}\label{eq53}
  u(x,t)=\la_3-\la_1-\la_2-2(\la_3-\la_2)\sn^2(\sqrt{\la_3-\la_1}\,
  (x-Vt),m),
\end{equation}
where
\begin{equation}\label{eq54}
  V=-2(\la_1+\la_2+\la_3),\qquad m=\frac{\la_3-\la_2}{\la_3-\la_1},
\end{equation}
and the parameters $\la_1\leq\la_2\leq\la_3$ are the zeroes of the third
degree polynomial
\begin{equation}\label{eq55}
  P(\la)=\prod_{i=1}^3(\la-\la_i)=\la^3-s_1\la^2+s_2\la-s_3
\end{equation}
defining the solution according to Eq.~(\ref{eq11}). Then
\begin{equation}\label{eq56}
  g(x,t)=\la-\mu(x,t),
\end{equation}
where $\mu(x,t)$ is connected with $u(x,t)$ by the relation
\begin{equation}\label{eq57}
  u=2\mu-(\la_1+\la_2+\la_3)=2\mu-s_1
\end{equation}
and depends only on the phase
\begin{equation}\label{eq58}
  \theta=x-Vt,\qquad \mu_{\theta}=2\sqrt{-P(\mu)}, \quad \la_2\leq\mu\leq\la_3.
\end{equation}
The wavelength of the solution (\ref{eq53}) is given by the expression
\begin{equation}\label{eq59}
  L=\frac12\oint\frac{\mathrm{d}\mu}{\sqrt{-P(\mu)}}
  =\int_{\la_2}^{\la_3}\frac{\mathrm{d}\mu}{\sqrt{-P(\mu)}}=\frac{2\K(m)}{\sqrt{\la_3-\la_1}},
\end{equation}
$\K(m)$ being the complete elliptic integral of the first kind.
In this one-phase case we can average over wavelength and the
Whitham velocities are easily expressed in terms of $L$ by
means of the formula
\begin{equation}\label{eq60}
  \Big\langle\frac1{g}\Big\rangle_{\la=\la_i}=\Big\langle\frac1{\la_i-\mu}\Big\rangle=
  \frac1{L}\int_0^L\frac{\mathrm{d} x}{\la_i-\mu}=\frac1{2L}\oint
  \frac{\mathrm{d}\mu}{(\la_i-\mu)\sqrt{-P(\mu)}}=-\frac2L\frac{\prt L}{\prt\la_i},
\end{equation}
so that we obtain
\begin{equation}\label{eq61}
  v_i=-2s_1+\frac{2L}{\prt L/\prt\la_l}=\left(1-\frac{L}{\prt_iL}\prt_i\right)V,
  \quad \prt_i\equiv\frac{\prt}{\prt\la_i},\quad i=1,2,3.
\end{equation}
Substitution of (\ref{eq59}) into this expression yields the well-known formulae
obtained by Whitham \cite{whitham65,whitham74}.
In the KdV equation case the sum in the right hand side of Eq.~(\ref{eq46}) reduces to a
single term with $\prt\mathcal{A}/\prt u=-1$. Taking into account that
$\sigma=-1$, $g=\la_i-\mu=(2\la_i-s_1-u)/2$, we arrive at
\begin{equation}\label{eq62}
  \frac{\prt\la_i}{\prt t}+v_i\frac{\prt\la_i}{\prt x}=
  \frac{L}{\prt L/\prt\la_i}\frac{\langle(2\la_i-s_1-u)R\rangle}
  {4\prod_{j\neq i}(\la_i-\la_j)}, \quad i=1,2,3.
\end{equation}
This is a compact expression for perturbed Whitham equations obtained in
\cite{MG95}.

Further calculation can be done after specification of the perturbation $R$.
For example, if we take
\begin{equation}\label{eq63}
  R=\eps u_{xx},
\end{equation}
($\eps$ is a small numeric parameter) what corresponds to taking into
account small viscosity in hydrodynamic applications of the KdV equation,
then we have
$$
u_{xx}=2\mu_{xx}=-2\frac{\mathrm{d} P}{\mathrm{d}\mu},\qquad 2\la_i-s_1-u=2(\la_i-\mu),
$$
so that
$$
\begin{array}{ll}
\langle(2\la_i-s_1-u)\eps u_{xx}\rangle & = -\frac{4\eps}L\oint
\frac{\la_i-\mu}{\sqrt{-P(\mu)}}\frac{\mathrm{d} P}{\mathrm{d}\mu}\mathrm{d}\mu=
\frac{8\eps}L\oint\sqrt{-P(\mu)}\mathrm{d}\mu \\ & =
\frac{16\eps}L\int_{\la_2}^{\la_3}\sqrt{-P(\mu)}\mathrm{d}\mu,
\end{array}
$$
and standard calculation of this elliptic integral yields
\begin{equation}\label{eq64}
\begin{split}
  \langle(2\la_i-s_1-u)\eps u_{xx}\rangle=
  \frac{32\eps}{15}\frac{(\la_3-\la_1)^3}{\K(m)}
  \big[(1-m+m^2)\E(m)\\-(1-m)(1-m/2)\K(m)\big],
  \end{split}
\end{equation}
$\E(m)$ being the complete elliptic integral of the second kind.
Substitution of (\ref{eq64}) into (\ref{eq62}) gives the perturbed
Whitham equations for the KdV-Burgers equation found first in
\cite{GP87,AN87}. Thus, the general formula (\ref{eq46}) reproduces
quite easily the already known results for the perturbed KdV equation.

\subsection{Nonlinear Schr\"odinger equation}

As the second example, let us consider the focusing NLS equation with linear damping
\begin{equation}\label{eq84}
  \mathrm{i}u_t+u_{xx}+2|u|^2u=-\mathrm{i}\eps u.
\end{equation}
(Similar results for the defocusing NLS equation will be presented
in the end of this Section.)
The unperturbed equation is usually associated in framework of the IST
method with the $2\times 2$ Zakharov--Shabat spectral problem \cite{ZS73}.
However, with the use of
Eqs.~(\ref{eq7}),(\ref{eq8}) it can be transformed into scalar form
(\ref{eq4},\ref{eq5}) with (see \cite{Alber,KK02})
\begin{equation}\label{eq85}
  \mathcal{A}=-\left(\la-\frac{\mathrm{i}u_x}{2u}\right)^2-|u|^2-
  \left(\frac{u_x}{2u}\right)_x,\qquad
  \mathcal{B}=2\la+\frac{\mathrm{i}u_x}u.
\end{equation}
Expansions of $\tilde{g}$ and $1/\tilde{g}$ are given by
\begin{equation}\label{eq86}
  \tilde{g}=1+\frac1{\la}\frac{\mathrm{i}u_x}{2u}-\frac1{\la^2}\left(\frac{|u|^2}2+
  \frac{u_{xx}}{4u}\right)-\frac1{\la^3}\left(\frac{\mathrm{i}u_{xxx}}{8u}-
  \frac{3\mathrm{i}u_xu^*}4\right)+\ldots,
\end{equation}
\begin{equation}\label{eq87}
\begin{split}
  \frac1{\tilde{g}}=1-\frac1{\la}\frac{\mathrm{i}u_x}{2u}+\frac1{\la^2}\left(\frac{|u|^2}2+
  \frac{u_{xx}}{4u}-\frac{u_x^2}{4u^2}\right)\\
  +\frac1{\la^3}\left(\frac{\mathrm{i}u_{xxx}}{8u}+
  \frac{\mathrm{i}u_xu^*}4-\frac{\mathrm{i}u_xu_{xx}}{4u^2}+
  \frac{\mathrm{i}u_x^3}{8u^3}\right)+\ldots,
  \end{split}
\end{equation}
and it is easy to check that they satisfy the identities
\begin{equation}\label{eq88}
  \frac{ \widehat{\delta }}{ \delta u}\frac1{\tilde{g}}=
  -\frac{1}{2\la^2}\left[ \frac{\prt\mathcal{A}}{\prt u}\tilde{g}
  -\frac{\prt}{\prt x}\left(\frac{\prt\mathcal{A}}
  {\prt u_{x}}\tilde{g}\right)
  +\frac{\prt^2}{\prt x^2}\left(\frac{\prt\mathcal{A}}
  {\prt u_{xx}}\tilde{g}\right)\right],
\end{equation}
\begin{equation}\label{eq89}
   \frac{ \widehat{\delta} }{ \delta u^*}\frac1{\tilde{g}}=
   \frac{u}{2\la^2}\,\tilde{g}.
\end{equation}
The generating function (\ref{eq25}) yields the conservation laws
in unusual form
\begin{equation}\label{eq90}
  \left(\frac{\mathrm{i}u_x}u\right)_t=-\left(2|u|^2+\frac{u_{xx}}u\right)_x,
\end{equation}
\begin{equation}\label{eq91}
  \left(\mathrm{i}|u|^2+\frac{\mathrm{i}u_{xx}}{2u}-\frac{\mathrm{i}u_x^2}{2u^2}\right)_t=
  \left(  \frac{u_xu_{xx}}{2u^2}- \frac{u_{xxx}}{2u}-2u_xu^*
  \right)_x, \ldots
\end{equation}
The one-phase periodic solution is parameterized by two pairs of
complex conjugated zeroes $\la_i$ of the polynomial (see
\cite{kamch2000,kamch90})
\begin{equation}\label{eq92}
  P(\la)=\prod_{i=1}^4(\la-\la_i)=\la^4-s_1\la^3+s_2\la^2
  -s_3\la+s_4,
\end{equation}
\begin{equation}\label{eq93}
  \la_1=\alpha+\mathrm{i}\gamma,\quad \la_2=\beta+\mathrm{i}\delta,\quad
  \la_3=\alpha-\mathrm{i}\gamma,\quad \la_4=\beta-\mathrm{i}\delta,
\end{equation}
so that the intensity of the wave is given by
\begin{equation}\label{eq94}
\begin{split}
  \nu=|u(x,t)|^2=\nu_3-(\nu_3-\nu_2)\sn^2(\sqrt{\nu_3-\nu_1}\,\theta,m)\\
  =(\gamma+\delta)^2-4\gamma\delta\,\sn^2(\sqrt{(\alpha-\beta)^2+
  (\gamma+\delta)^2}\,\theta,m),
  \end{split}
\end{equation}
where
\begin{equation}\label{eq95}
  m=\frac{\nu_3-\nu_2}{\nu_3-\nu_1}=\frac{(\la_1-\la_3)(\la_2-\la_4)}
  {(\la_1-\la_4)(\la_2-\la_3)}=\frac{4\gamma\delta}{(\alpha-\beta)^2+
  (\gamma+\delta)^2},
\end{equation}
$\nu_1,\,\nu_2,\,\nu_3$ are the zeroes of the resolvent $\mathcal{R}(\nu)$
of the polynomial $P(\la)$,
\begin{equation}\label{eq96}
  \begin{split}
  \nu_1=-\frac14(\la_1-\la_2+\la_3-\la_4)^2=-(\alpha-\beta)^2,\\
  \nu_2=-\frac14(\la_1-\la_2-\la_3+\la_4)^2=(\gamma-\delta)^2,\\
  \nu_3=-\frac14(\la_1+\la_2-\la_3-\la_4)^2=(\gamma+\delta)^2,
  \end{split}
\end{equation}
and $\nu$ satisfies the equation
\begin{equation}\label{eq97}
  \nu_{\theta}=2\sqrt{-\mathcal{R}(\nu)}, \qquad \mathcal{R}(\nu)=
  \prod_{i=1}^3(\nu-\nu_i),\quad \nu_2\leq\nu\leq\nu_3,
\end{equation}
$\theta$ being the phase of the wave
\begin{equation}\label{eq98}
  \theta=x-Vt,\qquad V=-s_1=-\sum_{i=1}^4\la_i.
\end{equation}

In this case there are two fields $u$ and $u^*$, and the
corresponding perturbations are equal to
\begin{equation}\label{eq99}
  R=-\eps u, \qquad R^*=-\eps u^*,
\end{equation}
where $\eps$ is a small constant parameter.
Easy calculation gives
$$
\frac{\prt\mathcal{A}}{\prt u}R+\frac{\prt\mathcal{A}}{\prt u_x}
\frac{\prt R}{\prt x}+\frac{\prt\mathcal{A}}{\prt u_{xx}}
\frac{\prt^2 R}{\prt x^2}+\frac{\prt\mathcal{A}}{\prt u^*}R^*=
2\eps|u|^2=2\eps\nu,
$$
so that we have to average the expression $2\eps\nu g$. Now, $g(x,t)=
\la-\mu(x,t)$, where $\mu$ satisfies the equation
\begin{equation}\label{eq100}
  \mu_{\theta}=2\sqrt{-P(\mu)},
\end{equation}
and can be expressed as a function of $\nu$ (see \cite{kamch97,kamch2000,kamch90}),
\begin{equation}\label{eq101}
  \begin{split}
  \mu(\nu)=\frac{s}4-\frac{q+\mathrm{i}\sqrt{-\mathcal{R}(\nu)}}{2\nu},\quad
  \nu_2\leq\nu\leq\nu_3,\\ s=s_1, \quad q=\sqrt{-\nu_1\nu_2\nu_3}.
  \end{split}
\end{equation}
Thus, we obtain
\begin{equation}\label{eq102}
\begin{split}
  \langle 2\eps\nu g\rangle=\frac{2\eps}L\int_0^{L}\nu(\la_i-\mu)\mathrm{d} x=
  \frac{\eps}L\oint\nu\left(\la_i-\frac{s}4+\frac{q+\mathrm{i}
  \sqrt{-\mathcal{R}(\nu)}}{2\nu}\right)\frac{\mathrm{d}\nu}{\sqrt{-\mathcal{R}(\nu)}}\\=
  \frac{2\eps}L\int_{\nu_2}^{\nu_3}\frac{(\la_i-s/4)\nu+q/2}
  {\sqrt{(\nu_3-\nu)(\nu-\nu_2)(\nu-\nu_1)}}\,\mathrm{d}\nu,
\end{split}
\end{equation}
where
\begin{equation}\label{eq103}
  L=\frac12\oint\frac{\mathrm{d}\mu}{\sqrt{-P(\mu)}}=\int_{\nu_2}^{\nu_3}
  \frac{\mathrm{d}\nu}{\sqrt{-\mathcal{R}(\nu)}}=\frac{2\K(m)}{\sqrt{\nu_3-\nu_1}}=
  \frac{2\K(m)}{\sqrt{(\la_1-\la_4)(\la_3-\la_2)}}.
\end{equation}
The integral in (\ref{eq102}) reduces to standard elliptic integrals,
\begin{equation}\label{eq104}
  \langle2\eps\nu g\rangle=\frac{2\eps}{L\sqrt{\nu_3-\nu_1}}
  \left\{\left[(2\la_i-s/2)\nu_1+q\right]\K(m)+(2\la_i-s/2)(\nu_3-\nu_1)
  \E(m)\right\}.
\end{equation}
The averaged values $\langle1/g\rangle$ and $\langle\mathcal{B}/g\rangle$
are calculated as in the previous cases of the KdV equation.
As a result, we arrive at the perturbed Whitham equations in the form
\begin{equation}\label{eq105}
  \frac{\prt\la_i}{\prt t}+v_i\frac{\prt\la_i}{\prt x}=
  -\eps\rho_i, \quad i=1,2,3,4,
\end{equation}
where Whitham velocities are given by the expression
\begin{equation}\label{eq106}
  v_i=-\frac{\langle\mathcal{B}/{g}\rangle}{\langle1/{g}\rangle}=
  -{s_1}+\frac{L}{\prt L/\prt\la_l}=\left(1-\frac{L}{\prt_iL}\prt_i\right)V,
  \quad  i=1,2,3,4,
\end{equation}
which reduces after substitution of (\ref{eq103}) to well-known formulae
\cite{FL86,pavlov87}, and functions $\rho_i$ in the right hand side of
Eq.~(\ref{eq105}) are equal to
\begin{equation}\label{eq107}
  \begin{split}
  \rho_1=-\frac{\left[(2\la_1-s/2)\nu_1+q\right]\K(m)+(2\la_1-s/2)(\nu_3-\nu_1)
  \E(m)}{(\la_1-\la_4)[(\la_1-\la_2)\K(m)+(\la_2-\la_3)\E(m)]},\\
  \rho_2=\frac{\left[(2\la_2-s/2)\nu_1+q\right]\K(m)+(2\la_2-s/2)(\nu_3-\nu_1)
  \E(m)}{(\la_2-\la_3)[(\la_1-\la_2)\K(m)-(\la_1-\la_4)\E(m)]},\\
  \rho_3=\frac{\left[(2\la_3-s/2)\nu_1+q\right]\K(m)+(2\la_3-s/2)(\nu_3-\nu_1)
  \E(m)}{(\la_2-\la_3)[(\la_3-\la_4)\K(m)-(\la_1-\la_4)\E(m)]},\\
  \rho_4=-\frac{\left[(2\la_4-s/2)\nu_1+q\right]\K(m)+(2\la_4-s/2)(\nu_3-\nu_1)
  \E(m)}{(\la_1-\la_4)[(\la_3-\la_4)\K(m)+(\la_2-\la_3)\E(m)]}.
  \end{split}
\end{equation}
Equations (\ref{eq106}) determine evolution of the Riemann invariants $\la_i,$
$i=1,2,3,4,$ due to combined action of modulation and perturbation effects.

For perturbed defocusing NLS equation
\begin{equation}\label{eq108}
  \mathrm{i}u_t+u_{xx}+2|u|^2u=-\mathrm{i}\eps u
\end{equation}
calculations are actually the same, so we present here only the final results.
The intensity of periodic wave is given by the formula
\begin{equation}\label{eq109}
\begin{split}
  \nu=|u(x,t)|^2=\bar{\nu}_1+(\bar{\nu}_2-\bar{\nu}_1)
  \sn^2(\sqrt{\bar{\nu}_2-\bar{\nu}_1}\,\theta,m)\\
  =\frac14(\la_1-\la_2-\la_3+\la_4)^2+(\la_2-\la_1)
(\la_4-\la_3)\,{\rm sn}^2\left(\sqrt{(\la_3-\la_1)(\la_4-\la_2)}\,
\theta,m\right)
  \end{split}
\end{equation}
where the zeroes $\la_i,$ $i=1,2,3,4$, of the fourth degree polynomial $P(\la)$
are real Riemann invariants, $\la_1\leq\la_2\leq\la_3\leq\la_4,$ connected with
$\bar{\nu}_i$, $i=1,2,3,$ by the formulae
\begin{equation}\label{eq110}
  \begin{split}
  \bar{\nu}_1=\frac14(\la_1-\la_2-\la_3+\la_4)^2,\quad
  \bar{\nu}_2=\frac14(\la_1-\la_2+\la_3-\la_4)^2,\\
  \bar{\nu}_3=\frac14(\la_1+\la_2-\la_3-\la_4)^2,
  \end{split}
\end{equation}
and
\begin{equation}\label{eq111}
  m=\frac{\bar{\nu}_2-\bar{\nu}_1}{\bar{\nu}_3-\bar{\nu}_1}
  =\frac{(\la_2-\la_1)(\la_4-\la_3)}
  {(\la_3-\la_1)(\la_4-\la_2)},
\end{equation}
\begin{equation}\label{eq112}
  \theta=x-Vt,\qquad V=-s_1=-\sum_{i=1}^4\la_i.
\end{equation}
The wavelength is equal to
\begin{equation}\label{eq114}
  L= \frac{2\K(m)}{\sqrt{\bar{\nu}_3-\bar{\nu}_1}}=
  \frac{2\K(m)}{\sqrt{(\la_4-\la_2)(\la_3-\la_1)}}.
\end{equation}
The perturbed Whitham equations are given by
\begin{equation}\label{eq115}
  \frac{\prt\la_i}{\prt t}+v_i\frac{\prt\la_i}{\prt x}=
  -\eps\rho_i, \quad i=1,2,3,4,
\end{equation}
where, as usual,
\begin{equation}\label{eq116}
  v_i=-\frac{\langle\mathcal{B}/{g}\rangle}{\langle1/{g}\rangle}=
  \left(1-\frac{L}{\prt_iL}\prt_i\right)V,
  \quad  i=1,2,3,4,
\end{equation}
and
\begin{equation}\label{eq117}
  \begin{split}
  \rho_1=-\frac{\left[(2\la_1-s/2)\bar{\nu}_3+q\right]\K(m)+(2\la_1-s/2)
  (\bar{\nu}_3-\bar{\nu}_1)
  \E(m)}{(\la_3-\la_1)[(\la_4-\la_1)\K(m)-(\la_4-\la_2)\E(m)]},\\
  \rho_2=-\frac{\left[(2\la_2-s/2)\bar{\nu}_3+q\right]\K(m)+(2\la_2-s/2)
  (\bar{\nu}_3-\bar{\nu}_1)
  \E(m)}{(\la_4-\la_2)[(\la_3-\la_2)\K(m)-(\la_3-\la_1)\E(m)]},\\
  \rho_3=-\frac{\left[(2\la_3-s/2)\bar{\nu}_3+q\right]\K(m)+(2\la_3-s/2)
  (\bar{\nu}_3-\bar{\nu}_1)
  \E(m)}{(\la_3-\la_1)[(\la_3-\la_2)\K(m)-(\la_4-\la_2)\E(m)]},\\
  \rho_4=-\frac{\left[(2\la_4-s/2)\bar{\nu}_3+q\right]\K(m)+(2\la_4-s/2)
  (\bar{\nu}_3-\bar{\nu}_1)
  \E(m)}{(\la_4-\la_2)[(\la_4-\la_1)\K(m)-(\la_3-\la_1)\E(m)]}.
  \end{split}
\end{equation}
Thus, we see that linear damping in the NLS equation leads to perturbed Whitham equations in
simple enough form suitable for applications.

It is easy to check that more general form of nonlinear damping,
$$
  \mathrm{i}u_t+u_{xx}\pm 2|u|^2u=-\mathrm{i}f(|u|^2)u,
$$
(with  a polynomial function $f(\nu)$)
as well as action of external potential $V(x)$ slowly depending on $x$,
$$
  \mathrm{i}u_t+u_{xx}\pm 2|u|^2u= V(x)u,
$$
 can be treated in the same way and arising
in this cases integrals can be reduced to standard elliptic ones, though the
formulae become more cumbersome and they will be presented elsewhere together with
appropriate application of the Whitham equations.

\section{Conclusion}

The examples considered in the preceding section show that the general
formulae (\ref{eq46}),(\ref{eq47}) provide an effective method of derivation
of perturbed Whitham equations. They can be applied to a number
of physically important equations integrable by the IST method with
$2\times 2$ matrix or second order scalar spectral problem. For one-phase
periodic solutions, the perturbation terms in Whitham equations are
expressed in the form of integrals which can be evaluated with the use of known
formulae obtained in derivation of the periodic solution under consideration.
Thus, the developed here theory permits one to treat effects of
perturbations in various physical problems related with modulated
nonlinear periodic waves. One such application has been done in the recent paper
\cite{AKK03} to evolution of modulated periodic waves in the Bose-Einstein
condensate under slow variation of the nonlinear coupling constant.

\section*{Acknowledgments}

I am grateful to F.Kh.~Abdullaev and V.V.~Konotop for useful discussions
and  to the staff of Centro de F\'{\i}sica da Mat\'eria
Condensada, Universidade de Lisboa, for kind hospitality.
The work there has been supported by the Senior NATO Fellowship.
I thanks also RFBR (Grant 01--01--00696) for partial support.

\setcounter{equation}{0}

\renewcommand{\theequation}{A.\arabic{equation}}


\section*{Appendix. Derivation of the identity (\ref{eq38})}

For simplicity we consider a strictly periodic case when
$u_m(x,t)$ are periodic functions of $x$ with period $L$. We start
from the statement that Eq.~(\ref{eq4}) is satisfied by the function
\begin{equation}\label{eq27}
  \psi^{\pm}=\sqrt{\tilde{g}}\exp\left(\pm \mathrm{i}\sqrt{-\sigma}
  \la^{r/2}\int^x\frac{\mathrm{d} x}{\tilde{g}}\right)
\end{equation}
provided $\tilde{g}$ satisfies Eq.~(\ref{eq21})
(see, e.g. \cite{kamch01}). These are
Baker-Akhiezer functions which can be considered as generalizations
of Bloch function solutions of the Schr\"odinger-like equation
(\ref{eq4}) with complex ``potential'' which is a periodic function
of $x$ with the period $L$. Then we can introduce formally a
``quasi-momentum'' $p$ as follows:
\begin{equation}\label{eq28}
  \psi^{\pm}=\exp(\pm \mathrm{i} px), \quad p=\frac{\sqrt{-\sigma}
  \la^{r/2}}{L}\int^L_0\frac{\mathrm{d} x}{\tilde{g}}.
\end{equation}
The ``quasi-momentum'' $p$ is a functional depending on $u_m,\, m=1,\ldots,N,$
and their space derivatives through the coefficients of the expansion
(\ref{eq26}). If we change $u_m$ a little, then $p$ will change a
little too, what defines the functional derivative
\begin{equation}\label{eq29}
  \frac{\delta p}{\delta u_m}=\frac{\sqrt{-\sigma}
  \la^{r/2}}{L}\frac{ \widehat{\delta} }{ \delta u_m}
  \left(\frac1{\tilde{g}}\right),
\end{equation}
where $\widehat{\delta} /\delta u_m$ is defined by (\ref{eq30}).

 On the other hand, this functional
derivative can be calculated directly in the following way. We take two
functions $u_m',\,u_m''$ with the same period $L$ and write equations for
corresponding Bloch functions,
\begin{equation}\label{eq31}
  \psi_{xx}'=\mathcal{A}(u_m')\psi',\quad \psi_{xx}''=\mathcal{A}(u_m'')\psi'',
\end{equation}
which give at once
\begin{equation}\label{eq32}
  \left[\psi_x'\psi''-\psi'\psi_x''\right]_0^L=
  \int_0^L\left( \mathcal{A}(u_m')- \mathcal{A}(u_m'')\right)\psi'\psi''\mathrm{d} x.
\end{equation}
Now, from Eq.~(\ref{eq27}) we find
\begin{equation}\label{eq33}
  \psi_x^{\pm}=\frac1{2\tilde{g}}\left(\tilde{g}_x\pm 2\mathrm{i}\sqrt{-\sigma}
  \la^{r/2}\right)\psi^{\pm},
\end{equation}
so that if $\psi^{\pm}$ has the property
\begin{equation}\label{eq34}
  \psi^{\pm}(x+L)=\mathrm{e}^{\pm \mathrm{i}pL}\psi^{\pm}(x),
\end{equation}
then $\psi^{\pm}_x$ has the same property,
\begin{equation}\label{eq35}
  \psi^{\pm}_x(x+L)=\mathrm{e}^{\pm \mathrm{i}pL}\psi^{\pm}_x(x),
\end{equation}
due to periodicity of $\tilde{g}$. Hence, for $\psi'=\psi'^-$ and
$\psi''=\psi''^+$ we find from (\ref{eq32}) that
\begin{equation}\label{eq36}
  \left(\mathrm{e}^{\mathrm{i}(p''-p')L}-1\right)\left(\psi_x'^-\psi''^+-
  \psi'^-\psi_x''^+\right)_{x=0}=
  \int_0^L\left(\mathcal{A}(u_m')-\mathcal{A}(u_m'')\right)\psi'^-\psi''^+\mathrm{d} x.
\end{equation}
Taking the difference $u_m''-u_m'=\delta u_m$ small, we obtain
$$
\psi_x'^-\psi''^+-\psi'^-\psi_x''^+\cong -2\mathrm{i}\sqrt{-\sigma}\la^{r/2}, \qquad
\mathrm{e}^{\mathrm{i}(p''-p')L}-1\cong \mathrm{i}\delta p\cdot L,
$$
and for the right hand side of Eq.~(\ref{eq36}) we get with the same accuracy
$$
-\int_0^L\left(\mathcal{A}(u_m'')-\mathcal{A}(u_m')\right)\tilde{g}dx\cong
-\int_0^L\left[\frac{\prt\mathcal{A}}{\prt u_m}\tilde{g}-
\frac{\prt}{\prt x}\left(\frac{\prt\mathcal{A}}{\prt u_{m,x}}\tilde{g}\right)+
\ldots\right]\delta u_m\mathrm{d} x,
$$
so that Eq.~(\ref{eq36}) yields
\begin{equation}\label{eq37}
  \frac{\delta p}{\delta u_m}=-\frac1{2\sqrt{-\sigma}L\la^{r/2}}
  \sum_{l=0}^{A_m}(-1)^l\frac{\prt^l}{\prt x^l}\left(\frac{\prt\mathcal{A}}
  {\prt u_m^{(l)}}\tilde{g}\right),
\end{equation}
where $A_m$ is the order of the highest derivative $u_m^{(A_m)}$  in $ \mathcal{A}$.
Then comparison of (\ref{eq28}) and (\ref{eq37}) yields the desired identity
(\ref{eq38}).

\end{document}